\documentstyle[12pt]{article}
\begin{document}
\renewcommand{\baselinestretch}{1.5}
\begin{flushright}
IP/BBSR/96-52 \\
hep-ph/9609397
\end{flushright}
\begin{center}
{\bf  CAN ONE STILL TALK OF A BROKEN QUANTUM CHROMODYNAMICS?} 
\end{center}
\vspace{1in}
\begin{center}
{\bf Afsar Abbas} \\
{Institute of Physics} \\
{Bhubaneswar-751005, India} \\
{e-mail : afsar@iopb.ernet.in}
\end{center}
\vspace{1.5in}
\begin{center}
{\bf Abstract}
\end{center}

The exact QCD is one of the cornerstones of particle physics.
Hence it would appear that there should exist no physical situation
where a broken version of QCD may have any relevance.
However, here I shall show that a spontaneously broken version of
QCD may have been significant in an early universe scenario.
How this can be made compatible with the exact QCD shall be
pointed out. This gives rise to a new cosmological dark matter
object (called "rungion").

\newpage

A proof of color confinement in QCD does not exist.However as there
has been no conclusive experimental evidence of free color the
confinement in QCD is largely believed to be true. In fact with time
the color confinement hypothesis has become one of the cornerstones
of the Standard Model (SM) of particle physics. As such a model such
as that of De Rujula, Giles and Jaffe (DGJ)
[1] and in which the local QCD is spontaneously 
broken, albeit slightly, would be met with scepticism today.
To anyone this spontaneously broken version of QCD and the 
currently popular unbroken 
version of QCD would be incompatible in any physically relevant
physical situation. In this letter I shall discuss a situation
pertaining to an early universe sceanrio wherein these two 
conflicting pictures may be brought together and in the end there
there would be nothing which conflicts with any experimental
information. In fact this makes predictions which would have 
tremendous phyical implications and which may be ultimately tested
in the cosmos and in the laboratory. In fact I shall point out
a situation where this prediction may already have been tested [2] !

Let us have a look at the DGJ model [1].
De Rujula, Giles and Jaffe  proposed a 
renormalizable spontaneously broken version of QCD. Color SU(3)
remains an exact global symmetry. The successful phenomenology
of color-singlet hadrons is altered very little and at the 
presently available energies the model is not inconsistent with
any experimental information. There is atleast one stable
hadron for every representation of color SU(3).
In their model the mass of the free quark is given as

\begin{equation}
M_{Q} = {1\over {2 \pi \alpha \mu}}
\end{equation}

and that of the free gluon is

\begin{equation}
M_{G} = \frac {3} {2} M_{Q}
\end{equation}

Here $\mu$  measures the breakdown of the local gauge symmetry and 
is zero in the standard QCD. $\alpha = .88 GeV^{-2}$.
For $\mu$ of 6 MeV the quark mass is 30 GeV while that of the
gluon would be 45 GeV. Hence the free quarks and gluons are 
very massive. Actually these quarks and gluons can be visualized
as complicated hadronic objects with a rather large size. For example
the 30 Gev quark will have a size of 3.9 fm while the 45 GeV gluon
would have a size of 4.5 fm. 
This DGJ model has already been
used by Bjorken and McLerran as a possible
means of explaining the puzzling Centauro events [2].

The currently very active ultrarelativistic heavy-ion collision programme
is being pursued with the hope of attaining the quark - gluon plasma
(QGP) phase of the hadronic matter. Earlier it was expected that
the hadron to QGP phase transition would take place at a single
temperature  $ T_{c}   \sim  150 - 200  $ MeV . 
Working in pure $ SU(3)_{c} $ theory
(where no quarks were present) it has been shown
that, free massless gluons which existed above $ T_{c} $ condense
through first order phase transition to a weakly interacting gas of
massive objects below $ T_{c} $ . There has been much discussion
as to whether the transition is first order or second order.

In contradiction to a single transition temperature as discussed
above, recently it has been becoming popular to talk of two distinct
QCD phase transition temperatures in QGP [3]. For example it has been
argued that the equilibrium of gluons takes place in time
 $ \tau_{g}   \sim   \frac{1}{2} $  fm/c while 
the production and equilibriation of quarks
takes place in $ \tau_{q}   \sim  2 $   fm/c  .  Hence one has a hotter
pure gluon plasma. It has been shown that the pure gluon plasma
transition temperature $ T_{g} \sim 400 $ MeV  and for the quark phase
$ T_{q}   \sim   250  $ MeV .
 
In the big - bang scenario as the universe cools it
goes through various phase transitions [4].
After the Electro - Weak symmetry is broken at temperature 
 $ \sim   200  $ GeV  , the particle content is $\gamma$,  $\nu_{e}$,
 $\bar{\nu_{e}}$,  $\nu_{\mu}$,  $\bar{\nu_{\mu}}$,  $\nu_{\tau}$,
 $\bar{\nu_{\tau}}$,  $e^{+}$ , $e^{-}$ ,  $\mu^{+}$ ,  $\mu^{-}$,
 $ u $,  $\bar{u}$,  $ d$,  $\bar{d}$,  and  gluons.
This remains so until the universe 
cools down to  a temperature where QCD permitted phase 
transition takes place. If the QCD phase ransition took place at one
single temperature then as per the standard model of QCD all hadrons
had better be color singlet.

However if QCD phase tarnsition takes place at two distinct 
temperatures then a very interesting possibilty arises.
We accept that as discussed above [3], there may be two distinct
QCD phase transition temperatures: one due to the pure gluon
plasma $ T_{g} \sim 400 MeV $ and the other one for the
quark phase at the lower temperature $ T_{q} \sim 250 MeV $.
Though they both arise from the standard QCD 
the exact nature of these QCD transitions is not clear.
The requirement that all the hadrons that 
constitute our universe today be color singlet, demands that
the QCD phase transition taking place last ( ie. at the lower 
teperature) be color confining as per the standard prescription.
But even in the standard prescription there are major uncertainties.
We do not know whether this phase transition is of 
first or second order . If this is the situation for the
standard QCD phase transition how can we say that the first
transition be exactly of the same nature as the second one?
In fact there is no a priori reason whatsoever for this
to be true. Below I shall discuss a scenario where this need not be
true and which leads to some possible interesting physics.

Let us assume that as the temperature of the early universe cools
down to $\sim 400 MeV$ the first hadronic phase transition takes 
place. We assume that this goes through the broken version of QCD
as suggested by De Rujula, Giles and Jaffe [1].Remember that this
temperature  ( 400 MeV ) is true for the pure gluon plasma phase 
only. Hence most ( or all ) gluons undergo first order
phase transition and condense to massive new objects called
"rungions" ( the meaning of this word shall be explained below ).
This newly created object would belong to the color octet
representaion. Rungions are what DGJ call the free gluon.
Though the DGJ model allows for objects belonging to all 
representations 3, 8, 10 etc 
due to the nature of this 400 MeV phase transition
the only new colored object created 
in this first phase transition would be the color octet rungion.
These rungions would be very heavy 
and stable and large sized ( compared to the other hadrons to arise 
later ) objects. We expect them to move around with 
non-relativistic velocities also.

There would be a freeze out and these rungions decouple from 
the rest of the particles.
There are very few free gluons present immediately after this
phase transition. However very soon due to the reaction 
$ qq \rightarrow gg $  the quarks " dress " themselves up. 
By the time the temperature drops to $ \sim  250 $ MeV, there are 
enough quarks and gluons  present to lead to the normal QCD phase 
transition to the standard hadronic matter consisting of color
singlet hadrons.
The existence of rungions is not incompatible with standard QCD as 
it would be very easy for two or more color octet rungions to come 
together to form a color singlet composite.

Irrespective of the details of the model, the very fact that  the
rungions were created and since they are heavy and stable, they will
contribute significantly to the total mass of the universe.
Hence they are a natural
candidate to explain the dark matter problem of cosmology [4].
The rungions are going to behave very differently from ordinary
hadrons as discussed in ref [1] and [2]. Various possibilities exist
but one thing is clear that the rungions would lead to some novel
and different physics and chemistry [1,2].

Here this new heavy dark matter candidate has been christened 
"rungion" taking the cue from the Hindoostanee word "rung" 
( pronounced as the 'rung' of a ladder ) meaning 'color'.
This is because what DGJ [1] call the
color octet gluon is actually the rungion in our picture.
This term will allow us to distinguish these new objects from
the massless gluons of the exact QCD.

These massive rungions become a relic species in the universe. To
get an estimate of the mass of the rungion let us assume that when
it decouples it is non relativistic : 
\begin{displaymath}
 T_{g} << m_{r}  
\end{displaymath}
Hence we obtain a conserved number N [4]
\begin{equation}
N = \frac{n}{s} = 0.145 (\frac{g}{g_{*}}) {( \frac{m_{r}}{T_{g}})}^
{ \frac {3} {2} }]
\exp(-\frac{m_{r}}{T_{g}})
\end{equation}

Where $ g_{*} $ is the number of species in equlibrium when the
rungion decouples. Taking $ \rho_{c} $ to be 
$ 1.06 \times 10^{4} h^{2}
eV-cm^{-3} $ and imposing the constraint 
\begin{equation}
(\Omega  h^{2})_{rungion}  \le  1
\end{equation}
 
If $ T_{g}  \sim  400  $ MeV  then we find that 
$ m_{r} \ge 46.2 $ GeV.  So our estimate is that the rungion
constituting the DM of the universe  is very heavy  $ \ge 46.2 $ GeV.
As shown above this mass is easily understood in the DGJ framework.

We know that there exist several candidates for the dark matter
[4]. There has been an extensive programme to identify the
dark matter. What has it to say regarding our prediction of
rungions of mass $\ge 46.2$ ?
Starkman, Gould, Esmailzadeh and Dimopoulos have studied in
detail [5] constraints on strongly interaction dark matter candidates
as arising from  several independent and diverse experiments
looking for the same. They allow for four windows, one of which is 
$ 3 GeV \le M_{DM} \le 10^4 GeV $. Hence our estimate of the
rungion dark matter mass is consistent with this experimental 
constraint. This is an important point in favour of the
arguments presented here.

\newpage

So what happens is this. The QCD phase transition at 400 MeV leads
to a slightly broken QCD with stable and heavy rungions
belonging to the octet representation of QCD.
This need not violate the colour confinement hypothesis of QCD
as we can very easily visualize a few rungions to come together
after the second QCD phase transition at 250 MeV to form a colour
singlet composite. Hence the existence of two different QCD phase
transition temperatures during the early universe allows for this
possibility. These rungions will contribute significantly to
the total mass of the universe and become a natural candidate
to explain the cosmological dark matter problem.
It is likely that rungions have already been detected as
having caused the Centaur events [2] (note that the Centaur objects
would now be color octet instead of color triplet).
It is important to note that the mass and the cross section range
of the rungions are very well permitted as per new studies on the
windows  on strongly interacting dark matter candidates based on 
global and comprehensive studies of various experimental 
constraints [5].


\newpage
{\bf References}
\vskip 50 mm

[1] A. De Rujula, R. C. Giles and R. L. Jaffe,\\
Phys. Rev. {\bf D17}, 285 (1978)\\

[2] J. D. Bjorken and L. D. McLerran,\\
Phys. Rev., {\bf D20}, 2353 (1979)\\ 

[3] E. Shuryak,\\
Phys. Rev. Lett. {\bf68} 3270 (1992)\\

[4] J. V. Narlikar,\\
'Introduction to Cosmology', Cambridge University Press (1993)\\

[5] G. D. Starkman, A. Gould, R. Esmailzadeh and S. Dimopoulos,\\
Phys. Rev. {\bf D41}, 3594 (1990)\\

\end{document}